\documentclass[10pt,conference,compsocconf]{IEEEtran}

\usepackage{epsfig}
\usepackage{amsmath}
\usepackage{amsfonts}
\usepackage{mathrsfs}
\usepackage{alltt}
\usepackage{url}
\usepackage{dblfloatfix}
\usepackage{balance}

\usepackage[english]{babel}

\begin{document}

% Page numbering for arXiv pre-print
\thispagestyle{plain}
\pagestyle{plain}

% Letter paper

\setlength{\pdfpageheight}{\paperheight}
\setlength{\pdfpagewidth}{\paperwidth}

% Identify paper as being presented at CompSys'2017
\IEEEoverridecommandlockouts

\IEEEpubid{\begin{minipage}{\textwidth}\ \\[12pt]
Presented at CompSys 2017 (\url{https://www.compsys.science/2017/home})
\end{minipage}}

\title{Providing A Compiler Technology-Based Alternative
For Big Data Application Infrastructures}
\author{%
% author names are typeset in 11pt, which is the default size in the author block
{K. F. D. Rietveld, H. A. G. Wijshoff}%
% add some space between author names and affils
\vspace{1.6mm}\\
\fontsize{10}{10}\selectfont\itshape
% 20080211 CAUSAL PRODUCTIONS
% separate superscript on following line from affiliation using narrow space
LIACS, Leiden University\\
Niels Bohrweg 1 \\
Leiden, The Netherlands\\
\fontsize{9}{9}\selectfont\ttfamily\upshape
\{krietvel,harryw\}@liacs.nl
}

\maketitle

\begin{abstract}
The unprecedented growth of data volumes has caused traditional approaches
to computing to be re-evaluated. This has started a transition towards the use
of very large-scale clusters of commodity hardware and has given rise to the
development of many new languages and paradigms for data processing and
analysis. In this paper, we propose a compiler technology-based alternative
to the development of many different Big Data application infrastructures.
Key to this approach is the development of a single intermediate
representation that enables the integration of compiler optimization and
query optimization, and the re-use of many traditional compiler techniques
for parallelization such as data distribution and loop scheduling. We show
how the single intermediate can act as a generic intermediate for Big Data
languages by mapping SQL and MapReduce onto this intermediate.
\end{abstract}

\begin{IEEEkeywords}
Big Data; Optimizing Compilers; Query Optimization; Program Transformation;
Parallelization; Distributed Computing
\end{IEEEkeywords}

\IEEEpeerreviewmaketitle

\section{Introduction}
During the past decade problem sizes and collected data volumes have been
increasing at an unprecedented rate. To be able to facilitate the processing
and analysis of such vast amounts of data, one has moved to the use of
large-scale clusters of commodity hardware. The larger the scale of such a
cluster, the larger the chance of hardware failure. Traditional approaches
to computing, such as database systems and high performance computing, do
generally not account for this, as they rely to be run on expensive,
high-end, high-availability hardware. Because of this, the emergence of
clusters of commodity hardware gave rise to the development of new software
processing frameworks with resilience to hardware failures built right in.

Given their reliance on massive amounts of commodity computer hardware
Google has fronted this movement, developing techniques such as
MapReduce~\cite{dean-2004}, BigTable~\cite{chang-2008}
and GFS~\cite{ghemawat-2003}. Outside of Google, several similar developments
occurred, most notably the Hadoop distributed processing framework, next to
an abundance of different data processing application frameworks. For
instance, several different approaches implementing distributed database
systems have been created, such as HBase, Hive and Cassandra, that all
bypass the relational model that has been traditionally used in database
systems.  Instead, these approaches rely on data structured in other manners
or on data that is not structured at all. The lack of a standard model for
such databases has led to a multitude of implementations, each with their
own advantages and disadvantages, different programming models and exposure
of low-level databases~\cite{meijer-2011}. Similarly, many different
frameworks for data processing and computation have been created partly
using similar techniques, including Pig, Tez, Mahout, Spark and Flink.
This in turn has led to efforts to make MapReduce more suitable for the
processing of relational data by re-envisioning MapReduce based on
tuples~\cite{ferrera-2012} and efforts to design a generic framework for the
optimization of declarative Big Data languages~\cite{borkar-2013}.
Finally, also variations of Hadoop have been described to make up for
deficiencies in Hadoop for particular use cases such as scientific computing
with commonly used scientific array-based data
formats~\cite{buck-2011,wang-2012,sehrish-2010}.

In this paper, we propose a compiler technology-based alternative to the
development of various different frameworks for Big Data Applications.
Similar to data mining applications, many Big Data applications run complex
queries on data sources to retrieve data and subsequently perform a
significant amount of further processing. For these computations to be
efficient they have to target both of these phases: data retrieval and data
processing.  This perspective forms the cornerstone of the alternative
proposed in this paper, which is to integrate compiler optimizations with
query optimization.  Or, put differently, compiler optimizations must become
aware of other storage levels that are in use besides memory and CPU cache
in a single computer.  By doing so, a single compiler intermediate is
created in which all optimization is carried out and from which executable
code is generated.

The use of this single intermediate solves deficiencies of data mining and
Big Data applications.  For instance, in data mining, data needs to be
restructured to suit the technique/analysis that one wants to apply. So,
data import and reformatting needs to take place before the computation can
ensue. Data to be imported can originate from different data sources,
including SQL databases. In case the data source is a transactional data
source that continuously updates, this expensive importing and reformatting
needs to take place continuously as well. Instead of reformatting the data,
one can also envision the computation to the reformatted -- i.e. bringing
the computation to the existing data format. A single intermediate
automatically enables the data import code and analysis/compute code to be
integrated. We refer to this as vertical integration: when data access code
is combined with the surrounding application code in the optimization
process, many more optimization opportunities can be unlocked. This has been
described for database applications~\cite{rietveld-2015-tods} and
Section~\ref{sec:vertical-integration} presents a concise introduction.

So, instead of using different frameworks, all problems can be expressed in
this single intermediate representation, allowing a single
``super''-optimizer to be employed. In fact, we propose to upgrade
traditional methods such that these can be used to successfully tackle Big
Data problems. This includes machine code generation, loop optimizations,
query optimizations (I/O optimizations) and techniques from parallel
compilers such as data distribution and decomposition. Furthermore, within
our approach additional techniques are available for the automatic
reformatting of data.  So, the compiler is equipped with tools to optimize
access to existing data, optimizing any processing done on this data and to
automatically generate new data structures to store re-formatted data for
optimized future processing.

In this paper, we highlight the possibility of the application (and re-use)
of traditional compiler techniques to contemporary Big Data problems.
A number of existing and newly developed techniques are discussed
that are to be combined into a single, powerful compiler framework for the
optimization of Big Data applications. Further research is required to
re-target compiler optimizations onto this framework, such that arbitrary
program codes and data distributions can be effectively optimized and
fault-tolerant code can be generated automatically.

This paper is organized as follows. Section~\ref{sec:vertical-integration}
discusses the integration of compiler optimizations and query optimization.
Section~\ref{sec:capabilities} discusses the capabilities of a single
intermediate representation, and parallelization and data access
optimization in particular. Section~\ref{sec:generic-intermediate} describes
how applications expressed in different paradigms can be expressed and
optimized within the single intermediate. Section~\ref{sec:implementation}
briefly discusses how the single intermediate is implemented.
Section~\ref{sec:conclusions} presents our conclusions.

\section{Integrating Compiler Optimizations And Query Optimization}
\label{sec:vertical-integration}
Within the vertical integration approach, the database queries that are
executed by the application are expressed and included in the same
intermediate representation as the application code, instead of sending
these queries to a DBMS for optimization and execution at run-time. This
representation allows traditional (loop) transformations from optimizing
compilers~\cite{padua-1986} to be applied on the queries (data access codes)
as well as allows for the (vertical) integration of data access codes with
generic application code, unlocking many more optimization opportunities.
Traditional analysis methods, such as Def-Use
analysis~\cite{allen-1976,kennedy-1981}, will detect and eliminate data
access of which the results are unused, or will detect related data accesses
that can be combined.  Separate loops that firstly read data from a database
and create a result set, and secondly process items in this result set, can
be automatically merged. In other words, the vertical integration approach
aims to optimize the database application as a whole.

\begin{figure}
\small
\begin{alltt}
\textbf{forelem} (i; i \(\in\) pA)
  \textbf{forelem} (j; j \(\in\) pB.id[A[i].b_id])
    \(\mathscr{R} = \mathscr{R} \cup\) (A[i].field, B[j].field)
\end{alltt}
\hrule
\begin{alltt}
\textbf{for} (int i = 0; i < A_len; i++)
  \textbf{for} (int j = 0; j < B_len; j++)
    \textbf{if} (A[i].b_id == B[j].id)
       R.append(A[i].field, B[j].field);
\end{alltt}
\hrule
\begin{alltt}
\textbf{for} (int j = 0; j < B_len; j++)
  hash.insert(B[j].id, B[j].field);
\textbf{for} (int i = 0; i < A_len; i++)
  \textbf{if} (hash.contains(A[i].b_id))
     R.add(A[i].field, hash.get(A[i].b_id));
\end{alltt}
\normalsize
\vskip -0.2cm
\caption{The figure demonstrates how from an initial specification in
the single intermediate
(top), different codes can be generated, for instance
using nested-loops join (middle) and an evaluation scheme using a hash
table (bottom).}
\label{fig:forelem-example}
\end{figure}

The main feature of the single intermediate relies on viewing data as being
stored as (multi)sets of tuples.  Within this framework a loop construct is
defined, \emph{forelem}, that defines an iteration over a (sub)set of these
multisets of tuples. The definition of the subset that is iterated is
performed using a feature called ``index sets'', that encapsulate how
exactly the iteration is carried out. The resulting loops are governed by
simple loop control, which allows these to be optimized using re-targeted
variants of traditional compiler optimizations. At a later compilation
stage, the compiler determines how to actually execute the iteration
specified by a \emph{forelem} loop and accompanied index set. This may be
done by nested loops iteration, but also through the use of hash functions
or tree-based indexes, leading to iteration patterns that are also employed
by database systems. See Figure~\ref{fig:forelem-example} for an example.

An important prerequisite for the above to be successful is the condition
that while using the single intermediate, queries can be optimized to a
level of performance that is at least competitive with ``traditional'' query
optimization technology. Query optimization using a single intermediate
representation has been described in~\cite{rietveld-2014-lcpc} and the
results show that using a single intermediate representation queries can be
optimized to an extent that surpasses the performance of state-of-the-art
database systems.

In fact, this approach to query optimization solves two problems experienced
by traditional optimizing compilers and query optimizations to make these
future-proof. One of the major aims of optimizing compilers has always been
to optimize CPU and cache utilization. Other storage levels, such as
disk-based storage or an intermediate flash layer, were just addressed in a
minor way. This is because the majority of the problem or data set to be
processed by a computational kernel typically fits main memory, either of a
single node or the combined memories of nodes in a cluster computer. It is
clear that this will not remain the case, given the rapidly growing problem
sizes and data set volumes. Therefore, it is of utmost importance that
compilers are made capable to optimize codes operating on such data sets.
Currently, compilers are inherently incapable of doing this, as optimization
for out-of-memory problems was never included in its aim. The integration of
techniques for query optimization as is done in a single intermediate
solves this incapability.

On the other hand, the goal of query optimization is to minimize the number
of disk I/O operations performed by execution plans for the retrieval of
data. Because the sizes of databases were traditionally significantly larger
than main memories of computers, the optimization objective was typically to
reduce the number of disk I/O operations. However, in the last decade the
size of computer main memories has increased significantly, resulting in a
shift of focus on disk I/O towards the exploitation of intrinsic internal
features of computer systems. This has led to query optimizers to start to
incorporate compiler techniques in order to maximize performance. Also at
this point, the integration of compiler optimization and query optimization
solves this problem.

Because both the power of compiler optimizations as well as query
optimization are encompassed in a single intermediate representation, it is
a solid foundation upon which to build a compiler for Big Data applications.
For a detailed description of vertical integration of database applications,
we refer the reader to~\cite{rietveld-2015-tods}. In the next section, we
discuss the capabilities of the framework with regard to parallelization and
their use in the optimization of Big Data Applications.

\section{Capabilities Of The Single\\Intermediate Representation}
\label{sec:capabilities}
In this section the main capabilities of the single intermediate
representation are described. This includes in particular techniques for
parallelization and optimization of data access codes. Subsequently, a
number of further optimizations are described such as data reformatting and
generation of highly efficient machine codes.

\subsection{Parallelization Techniques}
A \emph{forelem} loop that is defined in the single intermediate
representation is inherently parallel. However, to in order to come to an
efficient manner to execute such loops on a parallel computer, a suitable
distribution of the work load is needed. The approach that is chosen relies
on the techniques derived from the optimization of program code and data
distribution to map program codes onto parallel computers, see for
instance~\cite{kennedy-1998}. In the single intermediate representation, the
partitioning of data is being handled by special loop constructs which
express parallel execution coupled with data partitioning. This data
partitioning can be specified by an ``automatic'' partitioning of the
value range of one or more particular fields in the multiset model. Based on
the initial partitioning a mechanism for loop scheduling is selected, which
schedules iterations of parallel loops onto available processors. Finally,
all loops in the application are considered together to optimize the data
decomposition, leading to a final distribution of the data. The entire
approach relies on compiler transformations, making this approach very
generic allowing multiple data decompositions to be considered at compile
time.

\subsubsection{Data Partitioning}
To control how data is partitioned, generic loop transformations are used
among which loop blocking and orthogonalization. Using these
transformations, index sets or multisets can be divided into $N$ parts
corresponding to parallelization to $N$ processors. Consider a loop of the
following form, with a multiset \verb!A! containing tuples with fields
\verb!field1! and \verb!field2!:

\small
\begin{alltt}
\textbf{forelem} (i; i \(\in\) pA)
  SEQ;
\end{alltt}
\normalsize

\noindent
To create a \emph{direct} data partitioning, the loop blocking
transformation is used to split the index set \verb!pA! into $N$ partitions:
%%%%
\[\mathtt{pA} = \mathtt{p}_1{A} \cup \mathtt{p}_2{A} \cup \ldots \cup \mathtt{p}_N\mathtt{A}\]

\noindent
Subsequently, the example loop becomes:

\small
\begin{alltt}
\textbf{for} (k = 1; k <= N; k++)
  \textbf{forelem} (i; i \(\in\) p\(\sb{k}\)A)
    SEQ;
\end{alltt}
\normalsize

\noindent
and this loop is parallelized by replacing the \emph{for} loop with a
\emph{forall} loop, indicating that the outermost loop is parallelized.
Instead place of \emph{direct} data partitioning, also \emph{indirect} data
partitioning can be used. In this case, loop blocking is not done based on
the iterated index set, but on the value range of one of the multiset's
accessed fields. So, array \verb!A! is to be distributed into $N$ partitions
based on \emph{field1}. The notation \emph{A.field1} denotes the set of
values of the \emph{field1} found in all subscripts of \verb!A!. If
\verb!X = A.field1!, then
%%%%
\[\mathtt{X} = \mathtt{X}_1 \cup \mathtt{X}_2 \cup \ldots \cup \mathtt{X}_N\]

\noindent
is a partitioning of \verb!X! into $N$ segments. This gives rise to the
following loop:

\small
\begin{alltt}
\textbf{forall} (k = 1; k <= N; k++)
  \textbf{for} (l \(\in\) X\(\sb{k}\))
    \textbf{forelem} (i; i \(\in\) pA.field1[l])
      SEQ;
\end{alltt}
\normalsize

\noindent
In this parallelized loop nest a processor $P_k$ is responsible for
processing partition $\mathtt{X}_k$ of this partitioning and will execute
the original \emph{forelem} loop only for $i \in \mathtt{pA}, l \in
\mathtt{X}_k: \mathtt{A[i].field1} = l$. So, each processor will iterate the
values of the assigned partition $\mathtt{X}_k$ and execute the original
\emph{forelem} loop for each value.  Note that in this example of
\emph{indirect} data partitioning, a distribution of the data of array
\verb!A! is not made explicit in the intermediate representation of the
loop.

\subsubsection{Loop Scheduling}
Data partitioning has resulted in an initial distribution of data to
available processors. In loop scheduling, a parallel loop's iterations are
to be scheduled onto the available processors. Although the loop schedule
can be chosen to be in line of the data partitioning, loop scheduling is in
principle done independent of the data partitioning. With the subsequent
data decompositioning stage, all loops in the application are considered to
choose a final distribution of the data that minimizes communication between
different processors and nodes.

In the literature, many static and dynamic approaches to loop scheduling
have been described. A static loop schedule is determined entirely at
compile-time~\cite{cierniak-1995}.  Examples of dynamic scheduling
approaches are Guided Self-Scheduling (GSS)~\cite{polychronopoulos-1987},
Trapezoid Self-Scheduling~\cite{tzen-1993} and feedback guided dynamic loop
scheduling~\cite{bull-1998}.  The principle of these dynamic scheduling
techniques is that iterations of loops are scheduled at runtime.  Iterations
are allocated in groups called chunks. The process starts with a large chunk
size and this size gradually decreases with the course of execution.
Processors that finish their chunk earlier than other processors are
assigned a new smaller chunk.  By doing so, the work is better balanced in
case the cost for each loop iteration, or chunk, is not equal.

\subsubsection{Fault Tolerance}
The selection of a suitable loop scheduling technique is key to a successful
fault tolerant execution. Static loop schedules are simple and have no
overhead at all, but have the major disadvantage that there is no real
possibility for run-time changes. So, essentially, when one of the parallel
nodes fails it is not possible to offload work to another node and the
computation has to be restarted. With dynamic loop scheduling this is a
possibility, when a node fails remaining iterations scheduled for that node
(or these that have not been scheduled at all) can be scheduled to other
nodes. One can even take one step further and devise hybrid schemes, where
at a higher level dynamic loop scheduling is carried out and chunks of data
are executed according to a static schedule with no overhead. When a node
within the static group fails, only that chunk has to be computed on another
set of nodes, something the dynamic loop scheduler at a higher level will
take care of.

Loop scheduling thus serves two important purposes. Firstly, node failures
are taken care of by re-scheduling iterations assigned to that node to one
or more other nodes. Secondly, dynamic loop scheduling enables automatic
calibration of chunk size, allowing the code to automatically adapt to
different clusters and different compute node assignments.

\subsubsection{Data Distribution}
The selection and optimization of the data partitioning is followed by the
optimization of the data distribution. At this stage, all parallel loops in
the application are considered to choose the actual distribution of the
data. Different loops in the application might the accessing the same data
according to a different partitioning. During the selection of the data
distribution, these partitionings could be modified slightly to avoid
expensive data re-distributions from being carried out between these loops.
There might also be an initial distribution of the data present, so (part
of) the data is already distributed on a set of nodes. If this is the case,
this is taken into account during data distribution to optimize the
performance of the application.

It is possible that not all constraints will be resolved during the data
distribution phase. Because of this, a parallel loop might be performing
accesses to data that are not available locally. Such accesses are resolved
by remote communication to a processor that does have the necessary data
available. This communication can occur in many different ways, for instance
using TCP or UDP over Ethernet or MPI over InfiniBand.  Also if a
re-distribution of the data has to take place between loops extensive
communication has to take place. Therefore, in optimizing the final data
distribution, this communication should be minimized as much as possible.

To demonstrate how expensive data re-distributions can be avoided by the
compiler through the use of loop transformations, consider the following
example. Two adjacent loops are considered that each access the same
multiset and are parallelized based on \emph{different} fields of the tuples
in the multiset: \verb!X = A.field1! and \verb!X = A.field2!. These loops
compute the multiplicity of all values of \verb!field1! and \verb!field2!
respectively using an aggregate function:

\small
\begin{alltt}
\textbf{forall} (k = 1; k <= N; k++)
  \textbf{for} (l \(\in X\sb{k}\)) \{
    \textbf{forelem} (i; i \(\in\) pTable.field1[l])
      count\(\sb{1,k}\)[Table[i].field1]++;
    \textbf{forelem} (i; i \(\in\) pTable.distinct(field1))
      \(\mathscr{R}\sb{1,k} = \mathscr{R}\sb{1,k} \cup\) (Table[i].field1,
                       count\(\sb{1,k}\)[Table[i].field1])
  \}
...
\textbf{forall} (k = 1; k <= N; k++)
  \textbf{for} (l \(\in X\sb{k}\)) \{
    \textbf{forelem} (i; i \(\in\) pTable.field2[l])
      count\(\sb{2,k}\)[Table[i].field2]++;
    \textbf{forelem} (i; i \(\in\) pTable.distinct(field2))
      \(\mathscr{R}\sb{2,k} = \mathscr{R}\sb{2,k} \cup\) (Table[i].field2,
                       count\(\sb{2,k}\)[Table[i].field2])
  \}
\end{alltt}
\normalsize

\noindent
Even if $\mathtt{A.field1} \equiv \mathtt{A.field2}$ and the two
decompositions are the same, data partitioning conflicts will occur. This is
because a partitioning of \texttt{A} based on \texttt{field1} is not equal to a
partitioning of \texttt{A} on \texttt{field2}. The fact that the column contents
are equal does not imply the column contents are in the same order (the
columns are multisets).

To solve these conflicts, either \texttt{A} is not distributed for the first
loop or an expensive re-distribution of the data is performed between the
first and the second loop. Evidently, both are suboptimal solutions.
However, in this case a better solution can be found by performing loop
transformations such that these end up using the same data distribution.
This is done by exploiting the possibility to reorder the loops such that
the two parallelized loops computing the \texttt{count} aggregate are
consecutive to one another. This is possible because these loops do not have
a dependency on the other loops (the second \emph{forall} loops) in the code
fragment. The two outermost loops iterate the same bounds, allowing
application of the Loop Fusion transformation:

\small
\begin{alltt}
\textbf{forall} (k = 1; k <= N; k++)
  \textbf{for} (l \(\in X\sb{k}\)) \{
    \textbf{forelem} (i; i \(\in\) pTable.field1[l])
      count\(\sb{1,k}\)[Table[i].field1]++;
    \textbf{forelem} (i; i \(\in\) pTable.distinct(field1))
      \(\mathscr{R}\sb{1,k} = \mathscr{R}\sb{1,k} \cup\) (Table[i].field1,
                       count\(\sb{1,k}\)[Table[i].field1])
    \textbf{forelem} (i; i \(\in\) pTable.field2[l])
      count\(\sb{2,k}\)[Table[i].field2]++;
    \textbf{forelem} (i; i \(\in\) pTable.distinct(field2)))
      \(\mathscr{R}\sb{2,k} = \mathscr{R}\sb{2,k} \cup\) (Table[i].field2,
                       count\(\sb{2,k}\)[Table[i].field2])
  \}
\end{alltt}
\normalsize

\noindent
In the case that $\mathtt{Table.field1} \equiv \mathtt{Table.field2}$,
another series of statement reordering and Loop Fusion is possible in the
loop body resulting in:

\small
\begin{alltt}
\textbf{forall} (k = 1; k <= N; k++)
  \textbf{for} (l \(\in X\sb{k}\)) \{
    \textbf{forelem} (i; i \(\in\) pTable.field1[l]) \{
      count\(\sb{1,k}\)[Table[i].field1]++;
      count\(\sb{2,k}\)[Table[i].field2]++;
    \}
    \textbf{forelem} (i; i \(\in\) pTable.distinct(field1))
      \(\mathscr{R}\sb{1,k} = \mathscr{R}\sb{1,k} \cup\) (Table[i].field1,
                       count\(\sb{1,k}\)[Table[i].field1])
    \textbf{forelem} (i; i \(\in\) pTable.distinct(field2))
      \(\mathscr{R}\sb{2,k} = \mathscr{R}\sb{2,k} \cup\) (Table[i].field2,
                       count\(\sb{2,k}\)[Table[i].field2])
  \}
\end{alltt}
\normalsize

\noindent
Because the two counting loops are fused, the loops use the same
partitioning of \verb!X!. In other words, the loops use the same data
distribution and no data redistribution is necessary in between loops.

Although the interaction of the different transformations as illustrated is
rather powerful, it should be noted that we have only considered one
particular case of two consecutive \emph{forelem} loops. In general,
applications are not that simple and consist of many data access codes,
embedded or not embedded in application code, so, the complexity of these
interactions will grow exponentially. Strategies will have to be developed
to keep the optimization process manageable. These strategies will also need
to work with a representation of the data distribution and the communication
in the single intermediate.

\subsection{Data Access Optimization}
Another major feature of the single intermediate is the capability of
optimizing data access codes.  To optimize the queries that have been
expressed in the \emph{forelem} form, traditional loop transformations and
new transformations that are based on existing compiler transformations are
used. For instance, the loop interchange transformation is used to push any
conditions on data to outer loops to decrease the amount of data that needs
to be read as much as possible. After the loops have been ordered in an
optimized form, efficient code is generated to execute these loops.  In
\emph{forelem} loops the exact iteration order is encapsulated in the
``index set''. So, at this point the compiler will determine iteration
methods for these loops and generate appropriate code. An iteration method
may or may not involve the use of an additional index structure. As could be
seen in Figure~\ref{fig:forelem-example}, in the middle example an iteration
scheme is generated that does not use an index structure and visits the
entire multiset. The example at the bottom shows that the compiler employed
a hash-based index to generate a more efficient iteration pattern.

The compiler is equipped with optimizations that set up index sets and
optimize these index structures. For instance, through analysis the
compiler can deduce that some parts of multiset do not have to be indexed
because these parts will not be accessed and sometimes an index can be
generated in such a way that it can be used for more than one \emph{forelem}
loop. Note that the generation of such index structures happens at run-time
and these structures are only temporarily.
For more details on query optimization within  a single
intermediate, we refer the reader to~\cite{rietveld-2014-lcpc}.

Another major benefit of translating data access queries to the single
intermediate representation is that the compiler is enabled to also optimize
the data import part of data mining or Big Data applications. Because of
this, the way data import takes place can be better aligned with subsequent
processing. Furthermore, any subsequent processing can be integrated with
the code that actually retrieves the data. For instance, consider the
following code which retrieves student grades to compute the weighted average
from a database:

\small
\begin{alltt}
res = query("SELECT grade, weight FROM grades " +
            "WHERE studentID = \{0\}", studentID);
avg = 0.0;
\textbf{while} (r \(\in\) res) \{
  avg += r.grade * r.weight;
\}
\textbf{print} ("Average grade: \{0\}", avg);
\end{alltt}
\normalsize

\noindent
Within a single intermediate, the query would no longer be executed by a
separate database system, but instead be expanded as a series of loops.
Subsequently, the data access loop and the \emph{while} loop performing
further processing can be merged within a single loop, resulting in:

\small
\begin{alltt}
avg = 0.0;
\textbf{forelem} (i; i \(\in\) pGrades.studentID[studentID]) \{
  avg += Grades[i].grade * Grades[i].weight;
\}
\textbf{print} ("Average grade: \{0\}", avg);
\end{alltt}
\normalsize

\noindent
Note that such a transformation is not automatically possible using existing
techniques if the query and processing code are not present in a single
intermediate.

\subsection{Further Optimizations}
The single intermediate is equipped with numerous other capabilities for
performing optimization. In this section, Data Reformatting and Classic Code
Optimizations are two capabilities that will be highlighted.

\subsubsection{Data Reformatting}
\label{sec:data-reformatting}
Within the single intermediate, data that is operated on is represented as
tuples stored in multisets. This is only an intermediate representation and
thus does not imply that the data is physically stored as such. During the
code generation stage, the compiler determines a physical storage scheme for
the data. Data may be stored by simply storing the tuples as records in a
binary file. The compiler can also generate compressed column schemes
wherein a column that enumerates a range of values is not physically stored
in full, but rather a description of the value range is stored to be
reconstructed when the data is read. These optimizations are all possible
because the compiler has the code where data is read and written all under
control.

Besides controlling \emph{how} tuples are stored, the compiler also controls
the structure of the tuples themselves. In database applications, the tuples
often represent the \emph{schema} of a database, which has been set at
design time of the database and is not altered by query optimizers. Within
the single intermediate, such alterations can be done, again because all I/O
code is under control of the compiler. By analyzing the different data
access codes that are executed by the application in combination with the
subsequent data processing code and even run-time performance feedback, the
compiler will be able to optimize the data model of the application.

The scope by which such optimizations are possible depends on the
application. Some applications have to operate on already existing data. So,
similar to pre-existing data distributions, the compiler has to take the
existing format of the data into account. Reformatting all data for a small
optimization is prohibitively expensive. In such cases, the compiler will
hold off from performing such optimizations. However, if the data is going
to be processed multiple times in the future, it will pay off to store the
data in a different format. The compiler can in these cases automatically
generate code that will re-format the data during the first time of
processing so that subsequent runs of the computation will be significantly
faster. On the other hand, if the data to be processed has not yet been
collected, the compiler will generate a ``data import'' or ``data load''
code that will store the data in a suitable format for subsequent processing
when the data is collected. Naturally, if in the future other computations
are written that operate on the same data, the compiler can again decide to
generate another data reformatting code.

\subsubsection{Code Optimization}
After optimization at the level of the single intermediate has been
completed, machine code has to be generated. The compiler will take
advantage of the large body of existing compiler optimizations that are
essential for generating highly efficient machine code. These optimizations
include, among others, dead code elimination, common subexpression
detection, constant propagation, and register allocation to optimize the
sequence of instructions and control flow. Loops in the program code have to
be targeted in particular to optimize for re-use of cached data. To do so,
loop transformations are used that target loop nests as a while. By
modifying the order in which loop iterations are executed, cache data re-use
can be significantly improved leading to performance improvements of one or
more orders of magnitude. Loops can be further optimized by taking advantage
of vectorization techniques, which also opens to door to new computing
paradigms, such as Intel's Xeon Phi accelerator, that rely extensively on
vectorization units for acceleration.

\section{A Generic Big Data Intermediate}
\label{sec:generic-intermediate}
The single intermediate representation is also a \emph{single} intermediate
in the sense that it is a generic intermediate in which Big Data
applications can be expressed that were developed in different paradigms.
The novelty of this approach is that all optimization, including
optimization of data access or queries, optimization of data processing or
computational code and parallelization, takes place in a single
intermediate. This goes beyond approaches that translate Hadoop programs to
SQL such that storage engine features like indexing can be taken advantage
of, but that break down when arbitrary code is encountered~\cite{iu-2010},
and beyond approaches that propose a query compiler framework solely for the
optimization of queries, or data access codes~\cite{borkar-2013}.

As an illustration of the single intermediate, this section describes how
a MapReduce-like problem expressed in SQL can be expressed in the single
intermediate and can be subsequently expressed as a MapReduce problem. As
examples, two examples from the original MapReduce paper~\cite{dean-2004}
are considered. The discussion is concluded with an initial performance
evaluation of the MapReduce problem implemented in Hadoop and a parallel
code generated using the single intermediate representation.

The first example concerns URL access count. Consider logs
of web page requests, which are mapped to tuples $(url, 1)$. The reduction
operator is described in the paper as mapping $(url, list(values))$ to
$(url, total\_count)$. Considering a multiset \emph{access}, with a single
column containing the URLs, this computation can be described as the
following SQL query:

\small
\begin{verbatim}
SELECT url, COUNT(url) FROM access GROUP BY url
\end{verbatim}
\normalsize

\noindent
This query is expressed in the single intermediate representation as
follows:

\small
\begin{alltt}
\textbf{forelem} (i; i \(\in\) pAccess.distinct(url))
  \(\mathscr{G} = \mathscr{G} \cup\) (access[i].url)
\textbf{forelem} (i; i \(\in\) p\(\mathscr{G}\)) \{
  count = 0
  \textbf{forelem} (j; j \(\in\) pAccess.url[\(\mathscr{G}\)[i].url])
    count++
  \(\mathscr{R} = \mathscr{R} \cup\) (\(\mathscr{G}\)[i].url, count)
\}
\end{alltt}
\normalsize

\noindent
The compiler will first apply a number of initial transformations (Iteration
Space Expansion and Code Motion in this case) to enable parallelization, and
subsequently parallelize the loop using techniques described in the
previous section. One of the resulting codes of this parallelization
process is the following code fragment, where \verb!X = Access.url!:

\small
\begin{alltt}
count = 0
\textbf{forall} (k = 1; k <= N; k++) \{
  count\(\sb{k}\) = 0
  \textbf{for} (l \(\in\) X\(\sb{k}\))
    \textbf{forelem} (i; i \(\in\) pAccess.url[l])
      count\(\sb{k}\)[access[i].url]++
\}
\textbf{forelem} (i; i \(\in\) pAccess.distinct(url))
  \(\mathscr{R} = \mathscr{R} \cup\)(access[i].url,\(\sum\sb{k=1}\sp{N}\)count\(\sb{k}\)[access[i].url])
\end{alltt}
\normalsize

\noindent
Note that the resulting \emph{forelem} code bears similarity to a
MapReduce program. In fact, the first loop maps every row of \emph{access}
to an accumulation of the \verb!access[i].url! subscript of the \verb!count!
array. This could be represented as a tuple $(url, 1)$.  The second loop
iterates over all keys, which are all distinct URLs in \emph{access} and
retrieves the result of an aggregate function, in this case \verb!count!.

In general, two adjacent \emph{forelem} loops where the former loop stores values
in an array subscripted by a field of the array being iterated, and the
latter loop accesses elements of this array, can be written as a MapReduce
program. The \emph{map} function iterates the table that is iterated by the
former loop. This table is fragmented by a MapReduce framework, so that each
instance of the \emph{map} function processes a table fragment. Instead of
writing to a global array, \verb!emitIntermediate! is called. For the above
example, tuples \verb!(access[i].url, 1)! are generated, where the $1$ is a
dummy value, because it is not used.

The example code increments the value stored in the \verb!count! array for
every occurrence of a value \verb!access[i].url!. In the MapReduce program,
a pair will be generated for every \verb!access[i].url!. So, the reduction
function has to increment a counter for every occurrence of the same value
\verb!access[i].url!. Because a MapReduce framework will collect all pairs
for a unique key, the reduction function simply needs to count all values
for every unique key.

If the above example is written in MapReduce pseudocode similar to that used
in~\cite{dean-2004}, the program would be:

\small
\begin{verbatim}
map(key, value):
  # Assume value represents content of Access table
  access = value
  for a in access:
    emitIntermediate(a.url, 1)
reduce(key, values):
  count = 0
  for v in values:
    count++
  emit(key, count)
\end{verbatim}
\normalsize

\noindent
Imagine the above example performed an operation\linebreak
\verb!sum[Table[i].field1] += Table[i].field2! instead. In this case, a
MapReduce program will emit pairs \texttt{(Table[i].field1,
Table[i].field2)} (note that, the dummy $1$ is now replaced with
\texttt{Table[i].field2}). The summation is performed in the \emph{reduce}
function by summing the values for every unique key \texttt{Table[i].field1}.

\begin{figure*}[ht!]
\begin{center}
\includegraphics[scale=0.5]{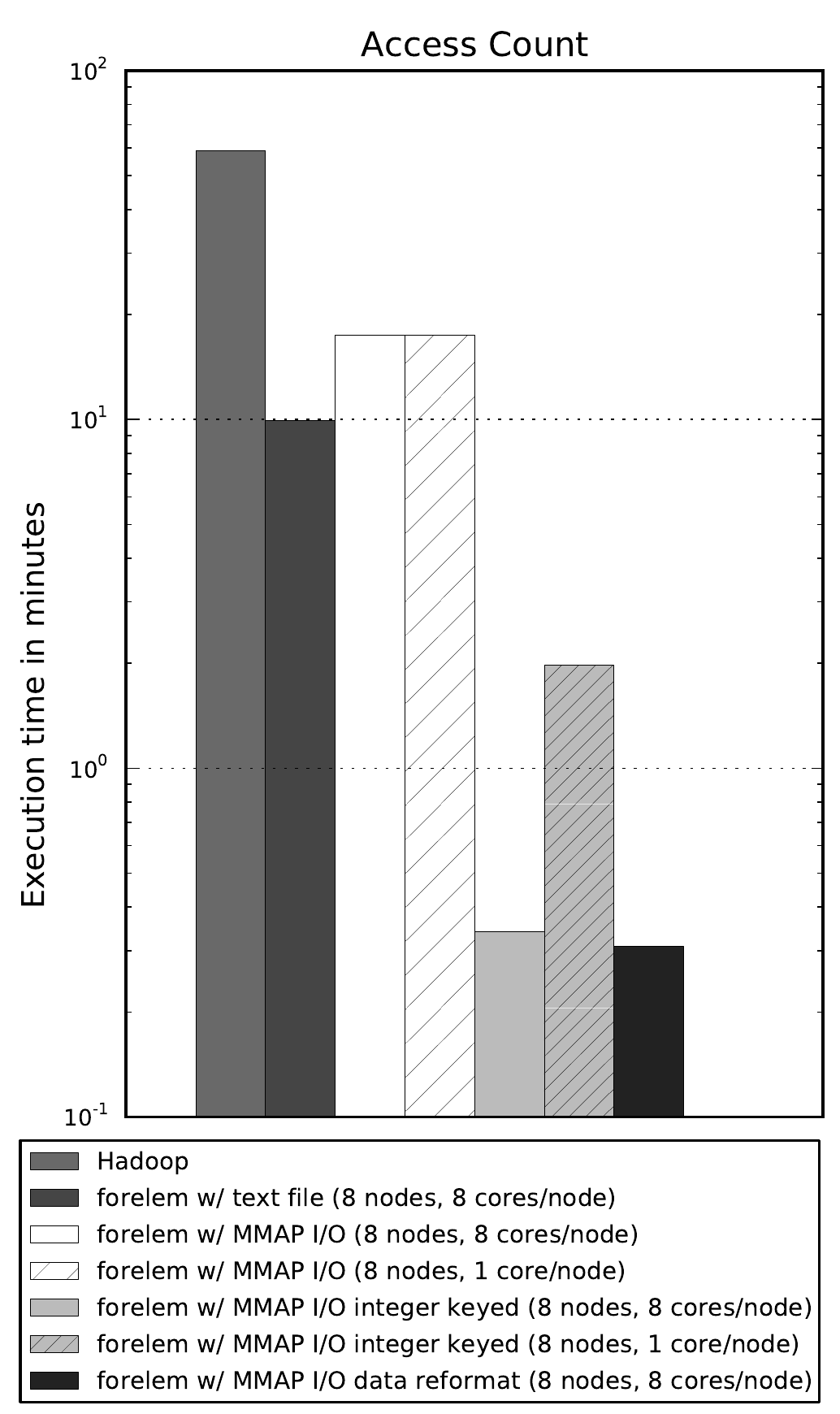}
\includegraphics[scale=0.5]{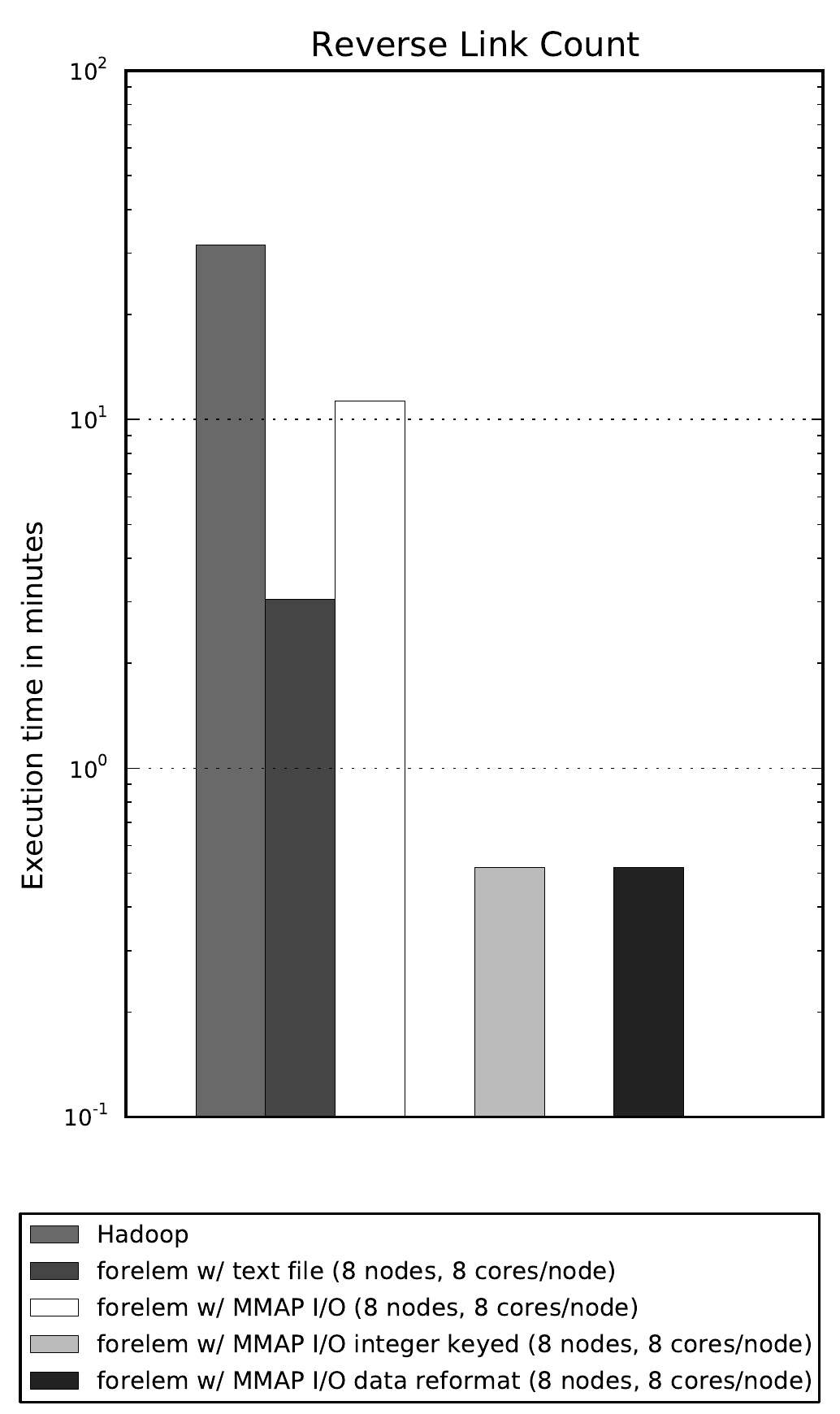}
\caption{Execution time in minutes for the Hadoop implementation and various
\emph{forelem} implementations of the two examples.}
\label{fig:experiments}
\end{center}
\end{figure*}

As a second example from the MapReduce paper, the Reverse Web-Link Graph is
considered. For each link from a source to a target page, a pair $(target,
source)$ is emitted. The original example reduces to a pair $(target,
list(source))$, which we will modify to reduce to a pair $(target,
\mathit{source\_count})$. To write a SQL query for this program, consider a table
\emph{links} that contains tuples $(source, target)$, which has been previously
filled, for example by parsing webpages \emph{source} and extracting all
links to target pages. The following two queries are defined:

\small
\begin{verbatim}
CREATE VIEW target_links AS
    SELECT DISTINCT target FROM links;
SELECT T.target, (SELECT COUNT(*) FROM links L
                  WHERE L.target = T.target)
FROM target_links T
\end{verbatim}
\normalsize

\newpage\noindent
which compute the number of incoming links to each registered target page.
Also this problem can be expressed in the single intermediate representation
and when parallelized with, for example, \texttt{X = Links.target}, the
resulting code is:

\small
\begin{alltt}
count = 0
\textbf{forall} (k = 1; k <= N; k++) \{
  count\(\sb{k}\) = 0
  \textbf{for} (l \(\in\) X\(\sb{k}\))
    \textbf{forelem} (i; i \(\in\) pLinks.target[l])
      count\(\sb{k}\)[links[i].target]++
\}
\textbf{forelem} (i; i \(\in\) pLinks.distinct(target))
  \(\mathscr{R} = \mathscr{R} \cup\) (links[i].target,
               \(\sum\sb{k=1}\sp{N}\)count\(\sb{k}\)[links[i].target])
\end{alltt}
\normalsize

\noindent
Note that also in this case a MapReduce code can be derived from this
intermediate representation using the technique as described above.

A number of experiments have been conducted with Hadoop and
implementations generated through a single intermediate representation of
the two described examples.  The experiments have been performed on the
DAS-4 cluster at Leiden University\footnote{Distributed ASCI
Supercomputer 4: \url{http://www.cs.vu.nl/das4/}}. The cluster nodes
each contain 8 processing cores, 48GB of main memory and 10 TB of local
storage in a software RAID0 configuration. The Hadoop experiments were
performed on a Hadoop cluster of 7 data nodes and one master node running
the task tracker. Using the single intermediate representation a C code has
been generated which uses MPI and OpenMP message
exchange and local parallelization. This implementation is also run on 7
nodes and one separate master node. The results are summarized in
Figure~\ref{fig:experiments}.  The experiments show that the generated
implementations realize an performance improvement of a factor 3 when the
same input data is used as is used by Hadoop, and up to a factor 120 if the
input data is available with an optimized layout.

As has been described in Section~\ref{sec:data-reformatting}, the single
intermediate representation is capable of automatically reformatting the
data layout of a program. As an example, the strings (URLs and hosts) in the
arrays have been replaced with integer keys. These integer keys are used to
subscript another array, which contains the string value for each key. In
fact, the data model has been made relational. This significantly improves
the performance, as indicated by the ``integer keyed'' experiments, which
implies that it is worthwhile to consider such data reformatting if this is
feasible in the context of the problem, for example when the data has not
yet been collected in a specific format. A final experiment has been done by
removing unused structure fields and column-wise storage of the data. These
data relayout operations can also be done automatically within the single
intermediate. A performance increase is not observed after performing this
relayout, possibly because it does not weigh up to the initial start up cost
of the MPI and OpenMP frameworks.

\newpage
\section{Implementation Of A Single\\Intermediate Representation}
\label{sec:implementation}
To be able to support different programming languages and database APIs,
a generic library was designed. This library is capable of creating
and manipulating \emph{forelem} loop nests, by representing these as
Abstract Syntax Trees (ASTs).
Also SQL statements can be parsed into an AST automatically.
On the AST, various analyses and
transformations can be applied, many of which are implementations of
traditional compiler (loop) transformations.
An abstract code generation interface is present in the
library to generate code from any \emph{forelem} AST.
For the vertical integration approach described in
Section~\ref{sec:vertical-integration}, query codes are integrated with, for
instance, a C++ code. In this case, the library is used from a prototype
Clang\footnote{\url{http://www.clang.org/}} compiler plugin. This plugin
scans a C/C++ AST for calls to database API and extracts the performed
operations, such as exact query strings that are requested to be executed.
The extracted information is passed to the library.  Transformations can
then be performed as an interplay between the C/C++ AST and the
\emph{forelem} AST created.  Finally, code in the C/C++ source code is
replaced with code generated using the library.

From a \emph{forelem} AST that contains parallel \emph{forall} loops, the
code generator incorporated in the library is capable of generating
parallelized code using the OpenMP and/or MPI frameworks. For contemporary
cluster computers that consist out of multi-core nodes, both frameworks are
used to achieve parallelization across nodes as well as across local CPU
cores. Similar to how other loop transformations are implemented in the
library, the library incorporates implementations of parallelization
transformations and logic for optimizing a data decomposition.  This data
decomposition support could consist out of a simple algorithm for
automatically generating a data decomposition based on a given AST, but also
out of a user interface to visualize the intermediate representation so that
the user can aid the framework in determining an efficient data
decomposition.

\section{Conclusions}
\label{sec:conclusions}
In this paper, we have presented a compiler technology-based alternative to
the development of various different frameworks for Big Data Applications.
This alternative is a single, generic intermediate representation and the
main features of this intermediate rely on the integration of compiler and
query optimizations and extensive parallelization capabilities. By using
this intermediate, a compiler is made capable of not only optimizing data
access and processing of data separately, but also to optimize data access
and subsequent processing together.

This compiler-based approach towards Big Data applications enables many
traditional compiler techniques to be re-used. For instance, we have
described how the work done on compiler-based parallelization of
applications, selection of data distributions and dynamic loop scheduling
can be fully re-used in the context of Big Data applications. By considering
multiple loops that access the same data within the same application the
compiler can use existing transformations to avoid expensive data
re-formatting between two loops. We have also demonstrated that the
intermediate is generic by showing how SQL queries can be expressed in the
single intermediate, be optimized and subsequently be expressed as a
MapReduce-like program. An initial experimental evaluation using two
examples also showed that by incorporating data reformatting capabilities in
the compiler, speedups can be attained of up to two orders of magnitude.

Further research is required to re-target compiler optimizations onto the
single intermediate such that arbitrary program codes and data distributions
can be effectively optimized. Using this intermediate, we aim to study the
feasibility and the advantages of applying the many existing techniques for
the optimization of parallel program codes to Big Data applications.
Furthermore, we would like to study the improvements in I/O optimizations
that can be achieved by integrating data processing and data
reformatting.
Finally, we plan to investigate novel hybrid loop scheduling
techniques that can be generated by a compiler and can gracefully handle
faults at run-time.

\IEEEtriggeratref{8}

\bibliographystyle{IEEEtran}
%\bibliography{references}


\end{document}